# Influence of Electrohydrodynamic on Droplet Stability in Leaky Dielectric Media

Saurabh Sharma[1], Franck Lefebvre[1], Mostafa S. Shadloo[1,2], Bertrand Lecordier[1*]
1 : 1INSA Rouen Normandie, Univ Rouen Normandie, CNRS, Normandie Univ, CORIA UMR 6614, 76801 Saint-Étienne-du-Rouvray, France
2 : Institut Universitaire de France, Rue Descartes, F-75231 Paris, France
* Correspondent author: bertrand.lecordier@coria.fr

ABSTRACT

This work examines the deformation dynamics of dielectric liquid droplet when exposed to a uniform electric field. The experimental investigation involves two-phase configurations, here as silicone oil droplets suspended in castor oil. The droplet dynamics including deformation, elongation, oscillation, and rotation are investigated over a range of electric field strengths using shadow-imaging. In parallel, fluorescent tracer particles were employed to perform three-dimensional Lagrangian Particle Tracking (3D-LPT) using the Shake-The-Box (STB) technique, allowing for a detailed characterization of the internal electrohydrodynamic flow within the droplet subjected to a uniform electric field. For silicone oil droplets in castor oil medium ($S/R > 1$), the droplets initially deform into oblate shape. At sufficiently high electric field strengths, the droplet undergoes an Electrohydrodynamic instability, aligning their axis at an angle to the direction of electric field due to the imbalance in induce electric torque. Upon further increasing the field strength, the droplets display oscillatory deformation before settling into a transiently stable configuration. The current study identifies and characterizes the distinct regimes of droplet behaviour, from initial deformation at low electric fields to oscillatory behaviour at high field strengths depending upon the relative dielectric and fluid properties of the liquid phases.

## 1. Introduction

Droplet electrohydrodynamics (EHD) flow finds its application in variety of physical phenomena that play a crucial role in processes such as electro-coalescence, electro-separation, and electro-spraying etc, as well as in enhancing mixing and transport in microfluidic devices. For an immiscible liquid-liquid interface subjected to a uniform external electric field, differences in the physical and electrical properties of the two liquids lead to a discontinuity in the normal component of the electric field across the interface. This discontinuity generates interfacial Maxwell stresses, which can deform the interface and induce electrohydrodynamic flows. (Vlahovska, P. M. 2019, Abbasi et al., 2020).

Numerous theoretical studies have emphasized the deformation behavior of droplets in the presence of an external electric field. In his classical electrostatic model, Garton and Krasucki (1964) proposed that for perfectly dielectric liquids, the resulting electric stress vector acts solely in the direction normal to the interface. In the perfectly dielectric model (PDM), the only normal component of the stress vector at the interface is counterbalanced by the interfacial tension, causing the droplet to elongate along the direction of the electric field, resulting in a prolate shape. However, several experimental studies indicate that the presence of an electric field can lead to two distinct regimes of droplet deformation, one elongation along the field direction (prolate deformation) and flattening orthogonal to it (oblate deformation) (Karp et al. 2024, Salipante & Vlahovska, 2010). In practical applications, no liquid behaves as a perfect dielectric. The presence of trace impurities imparts a finite, though non-zero, electrical conductivity. Such liquids are commonly referred to as leaky dielectrics. When a liquid droplet is immersed in another immiscible liquid, the application of an electric field across the curved interface of these two weakly conducting fluids generates electric stresses that can induce sustained fluid motion (Vlahovska, P. M. ,2019).

In 1966, G. I. Taylor (1966) explained this behavior through the leaky dielectric model (LDM), which describes how droplets deform under an electric field. This model agrees well with the experimental observations under moderate electric field strengths. However, when the applied electric field exceeds a critical value ($E_c$), discrepancies begin to emerge between the experimental results and the predictions made by the asymptotic theory (Karp et al. 2024).

In a uniform, weak DC electric field, the electric and velocity fields around an isolated, neutrally buoyant leaky dielectric droplet at very low Reynolds number exhibit azimuthal symmetry about the axis of the applied field (Karp et al. 2024). The equilibrium configuration of the droplet results from a balance between the electric stresses and the opposing hydrodynamic stresses and interfacial tension forces on its interface, and this behavior can be characterized in terms of the conductivity ratio (R) and permittivity ratio (S) of the two fluids. Under small electric fields, this balance yields a stable, steady-state droplet shape (Karp et al. 2024). However, when the applied field strength exceeds a critical threshold, the EHD instabilities develop, leading to continuous deformation of the droplet until it ultimately fragments into smaller droplets (Brosseau & Vlahovska, 2017). In the experiments of Salipante and Vlahovska (2010), it was reported that at stronger fields, drops exhibit Quincke-like rotation. It is driven by charge convection, orientation angle, and rotation rate and that too influenced by the viscosity ratio. Moreover, in their extended work, Salipante and Vlahovska (2013) reported that droplets in DC electric fields can show chaotic rotational dynamics, explained by polarization anisotropy and charge convection effects. These instabilities arise from an imbalance in the induced electric torque (Salipante & Vlahovska, 2013). Despite its significance, the fundamental physics of this phenomenon remain insufficiently understood, with only a limited number of studies investigating droplet dynamics at high electric field strengths, particularly within the oblate regime.

The present study investigates the electrohydrodynamic instability of liquid droplets subjected to an external electric field. Prior research has shown that a silicone oil droplet immersed in castor oil (S/R > 1) initially undergoes oblate deformation when exposed to the field (Karp et al. 2024). At subcritical field strengths, the droplet reaches a steady deformation state. However, once the applied field surpasses a critical threshold, the droplet exhibits Quincke-like rotation accompanied by deformations as well as alignment of droplet axis at an angle relative to the field direction.

## 2. Experimental Setup and Procedure

A custom 3D-printed cuvette was used, with its two sides covered by polymethylmethacrylate to enable droplet visualization. Two 1mm thick copper electrodes of size 4 × 8 cm each, were placed 20 mm far from each another, and a DC high-voltage power supply (Spellman SL100) was employed to generate the constant/uniform electric field between the two electrodes. A droplet of approximately 2.5 mm in diameter was manually pipetted into the center of the chamber, sufficiently far from both the electrode boundaries and the free surface to minimize wall and surface effects (Fig. 1a). In the current study, the droplet phase has been chosen of Silicon oil and surrounding medium is of castor oil, the measured properties of the oils are given in the Table 1. Throughout the investigation the applied electric field was kept between 3-6 kV/cm, with the voltage rise time of the power source measured to be less than 0.1s which is much lesser than the droplet response time to the electric field.

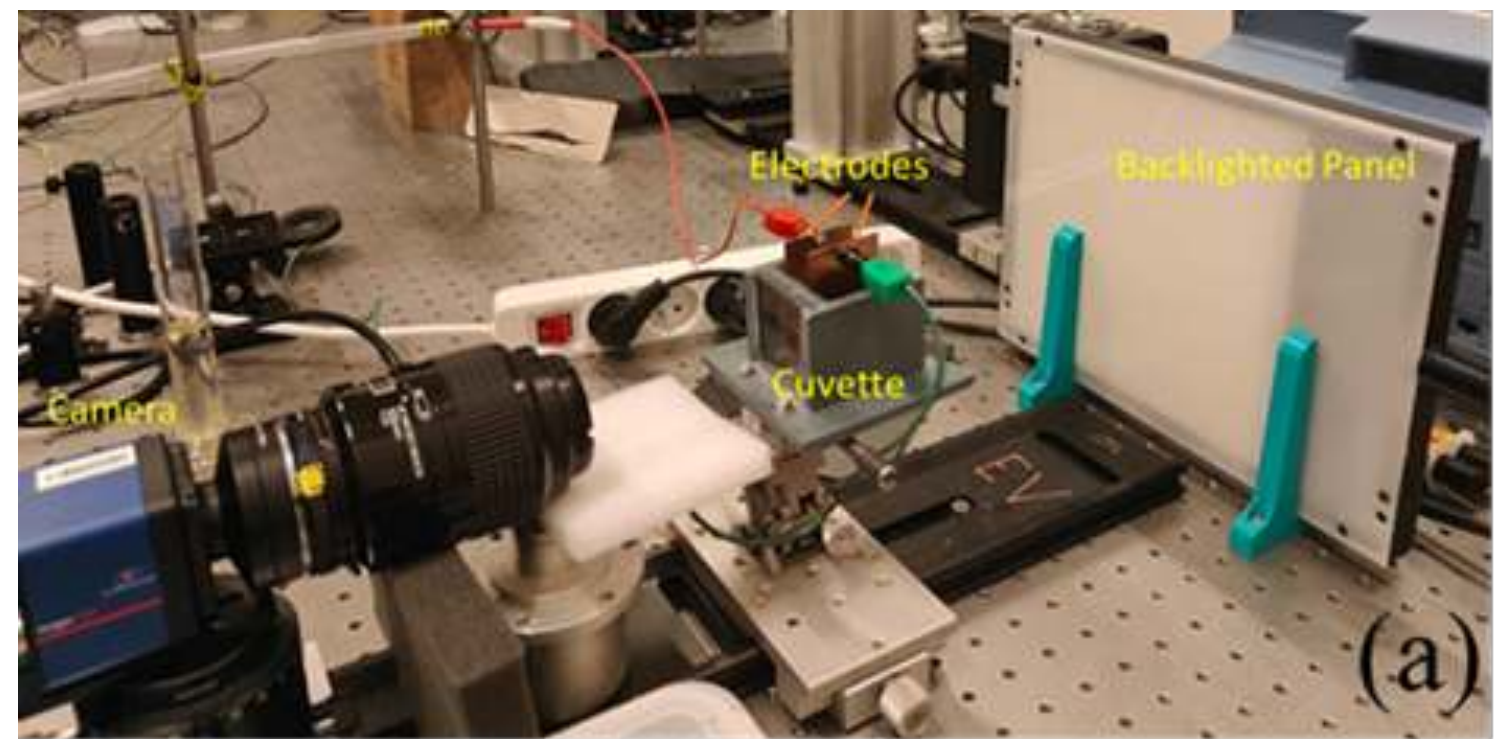

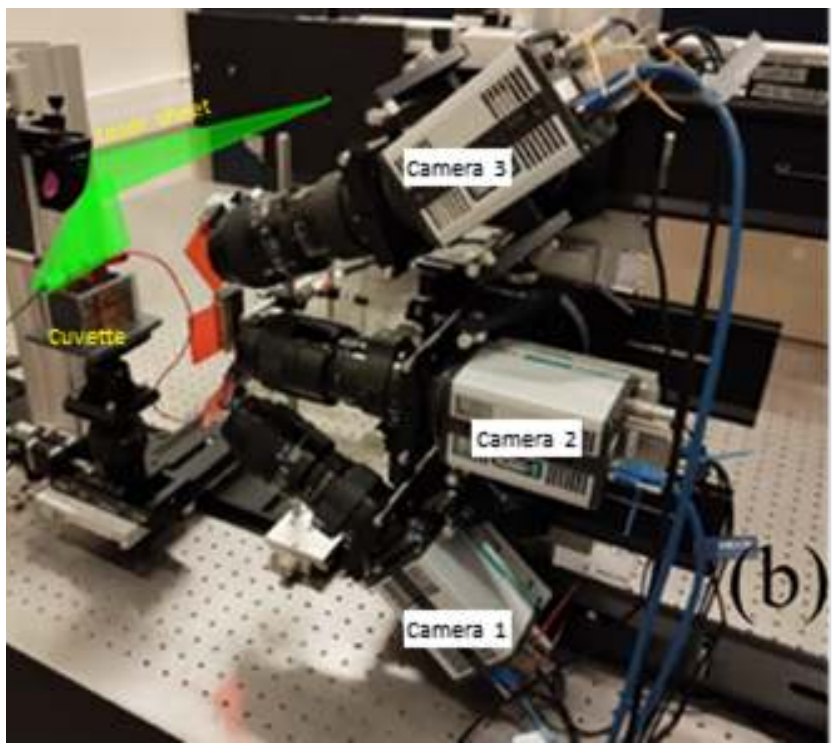

**Figure 1.** Experimental setup showing the image acquisition system for (a) Shadow imaging technique (b) 3D Lagrangian Particle Tracking technique.

***High resolution shadow imaging***

The transient dynamics of the droplet were recorded using a CCD camera (Imager CX3, LaVision, 5501×4648, ~25 MP resolution) operating at 50 frames per second (Fig. 1a). The chamber was aligned such that the applied field was oriented from left to right in the captured images. Image acquisition continued until the droplet reached equilibrium, typically ranging from 15–30 seconds depending on the applied field strength.

Image sequences were analyzed using ImageJ software along with in-house developed Python scripts. A pixel intensity threshold was applied to each image stack to extract the droplet interface. The processed images were then converted to binary, with the enclosed region filled in white. The image resolution was kept to 207 pixels/mm.

**Table 1.** Characteristics of the working fluids used in this study

| Property | Silicone oil | Castor oil |
|---|---|---|
| Density, ρ (kg/m$^3$) | 970 | 961 |
| Viscosity, $\mu$ (Pa.s) | 0.50 | 0.78 |
| Electrical permittivity, ε (F/s) | $2.83\times10^{-11}$ | $4.16\times10^{-11}$ |
| Electrical conductivity, σ (S/m) | $8.71\times10^{-13}$ | $3.01\times10^{-11}$ |
| Refractive index, n | 1.41 | 1.47 |

*The interfacial tension of the silicone-castor oil system is approximately 4 mN/m.

*Volumetric particle-tracking*

Furthermore, three-dimensional electrohydrodynamic flows both inside and around the droplet were investigated using a volumetric particle-tracking technique (Fig. 1b). The droplet and the surrounding liquids were seeded with 20μm Rhodamine-B coated glass tracer particles and illuminated by a 532 nm Nd:YAG laser (Quantel BRIO, France). Through a combination of optical elements, including a slit to maintain approximately uniform volume illumination, a thick laser sheet of about 1 cm was generated, illuminating the entire droplet volume as well as the surrounding region. Three synchronized cameras (HiSense Zyla, Dantec) were positioned in-line in front of the cuvette, each fitted with a high-resolution CMOS sensor and a 105 mm macro lens (Nikon Macro, Japan). However, owing to the small size of the region of interest (including the droplet and the surrounding medium) as well as optical constraints imposed by the presence of the metallic electrodes, the Scheimpflug principle was employed to compensate for camera inclination and maintain sharp focus across the measurement volume. The field of view of each camera was approximately 25 × 21 mm, enabling simultaneous visualization of the flow both inside the droplet and in the surrounding liquid. Image acquisition was conducted using DynamicStudio (Dantec Dynamics V8), which provided synchronization between the cameras and laser illumination system. The acquired image sequences were subsequently exported for post-processing and analysis in DaVis 11 (LaVision). Volumetric calibration and self-calibration procedures were performed to reconstruct three-dimensional particle positions and velocity fields using the Shake-the-Box (STB) Lagrangian Particle Tracking (LPT) method. Images were acquired at a rate of 20 Hz, which was found to be sufficient for accurately resolving the flow dynamics associated with the physical phenomena under investigation.

The reconstruction of the 3D volume from the acquired images from the 3 cameras was carried out using the polynomial camera mapping approach where the 3D coordinates were created with the help of 2D coordinates from the acquired images. A 3-D particle trajectories obtained from Shake-the-Box (STB) processing were subsequently used to reconstruct the continuous velocity field using the fine scale reconstruction VIC# (Vortex in Cell sharp) method implemented in DaVis. The VIC# procedure reconstructs a continuous velocity field on a regular Cartesian grid from Lagrangian particle tracks using a kernel-based and regularized iterative interpolation scheme that ensures spatial consistency of the velocity field.

For flow conditions, the velocity field was further averaged over the complete acquisition sequence (596 frames) to improve convergence and reduce random measurement noise. The fine-scale reconstruction employed an isotropic grid spacing of approximately 0.5mm (corresponding to 50 voxels per grid unit) yielding approximately 74,880 total vector nodes. The VIC# optimization was regularized using a denoising factor of 0.001 over 40 iterations, utilizing a time projection length of ±1 time step. The resulting field was used for the analysis of the electrohydrodynamic circulation inside and outside the droplet.

## 3. Results and Discussion

The present study investigates the oblate deformation of a droplet subjected to relatively high electric field strengths. To achieve oblate deformation, a silicon oil droplet suspended in a castor oil medium was considered, as this fluid system has S/R ratio greater than unity, which favors oblate shape deformation. The droplet behavior was examined under uniform electric fields of varying strengths, $E_0$ = 4.5, 4.75, 5.25, and 5.75 kV/cm. For the present system, the electric capillary number, defined as $Ca_E = (\epsilon_o E_o^2 a)/\gamma$ is varying from 2.6-4.3. The electric field strength ($E_0$) was determined by dividing the potential difference applied across the electrodes by the distance separating them, assuming a uniform electric field. This range of electric field strengths was chosen because droplets exhibit stable oblate deformation below it, whereas higher field strengths lead to droplet breakup.

The present study examines the deformation dynamics of a dielectric droplet subjected to a uniform DC electric field using a shadow imaging technique. As it can be seen from Figure 2 which represents high-resolution camera images of a dielectric droplet undergoing deformation under a uniform DC electric field.

As the degree of droplet deformation can be more conveniently quantified by analysing the variation of the droplet aspect ratio as a function of time. Therefore, Figure 3 shows the temporal evolution of the aspect ratio (AR) and the orientation angle (OA) of the deformed droplet for different electric field strength. The orientation angle is measured as the angle formed between the major axis of the droplet and the axis perpendicular to the applied electric field. The aspect ratio was determined by fitting a bounding rectangle around the deformed droplet, with the rectangle oriented along the droplet's principal axis of inclination.

The results indicate that the droplet's deformation dynamics exhibits distinct behaviors based on the variation of aspect ratio; such as sharp increase (Regime I), gradual increase or near invariance (Regime II), decrease (Regime III), equilibrium state (Regime IV), or a single cycle of down-upturn/oscillation (Regime V), which have been classified into separate regimes for better understanding.

At $E_0 = 4.5$ kV/cm (Fig. 3a), the aspect ratio (AR) of the deformed droplet initially shows a sharp rise (Regime I), followed by a slower, gradual increase (Regime II). This behavior reflects the droplet's rapid response to electric stresses. However, the rate of deformation estimated by variation of aspect ratio, decreases rapidly after attaining a peak. For the present case (S/R > 1), the induce stresses push fluid from pole to the equator and resulting into flattening of the droplet into an oblate shape. This increases the local curvature at the equator, thereby reducing the radius of curvature (Fig. 3a) Since the interfacial restoring pressure (Laplace pressure) scales as γ/a, with γ the interfacial tension and a is the radius of curvature, the growing interfacial tension force increasingly resists further deformation. Thus, the system tends toward a quasi-steady equilibrium state, where electrostatic stresses balance the surface tension force. In addition, as charge distribution reaches steady state, the associated hydrodynamic flows decay, further slowing the deformation.

Furthermore, it is evident from Fig. 3a that, simultaneously, with deformation, the droplet also begins to tilt with respect to the electric field direction. When the conductivity and permittivity ratios are such that (S/R > 1), the induced dipole moment aligns opposite to the applied field (Salipante & Vlahovska, 2010), the system becomes unstable. Above a critical field strength, the droplet undergoes spontaneous rotation. If the dipole tilts slightly, the torque exerted by the external field ($\tau_e = E_o \times P$) amplifies the rotation rather than restoring it. However, the tilt of the droplet is the instability arise due to rotation of the droplet.

Furthermore, the droplet aspect ratio exhibits a characteristic peak at the onset of rotation. Prior to rotation, the applied electric field generates axisymmetric electric stresses that strongly responsible for the stretching of the droplet normal to the field direction (flattening along the direction of field), thereby, increasing the aspect ratio. Once the instability is triggered, the induced surface charges rotate around the droplet, redistributing interfacial stresses. Consequently, the aspect ratio decreases after reaching its maximum value (regime III), reflecting the competition between electrohydrodynamic rotation and capillary relaxation. Thus, titling of droplet provides an additional mechanism, alongside with interfacial tension, that resists or arrests further axisymmetric deformation in one axial direction.

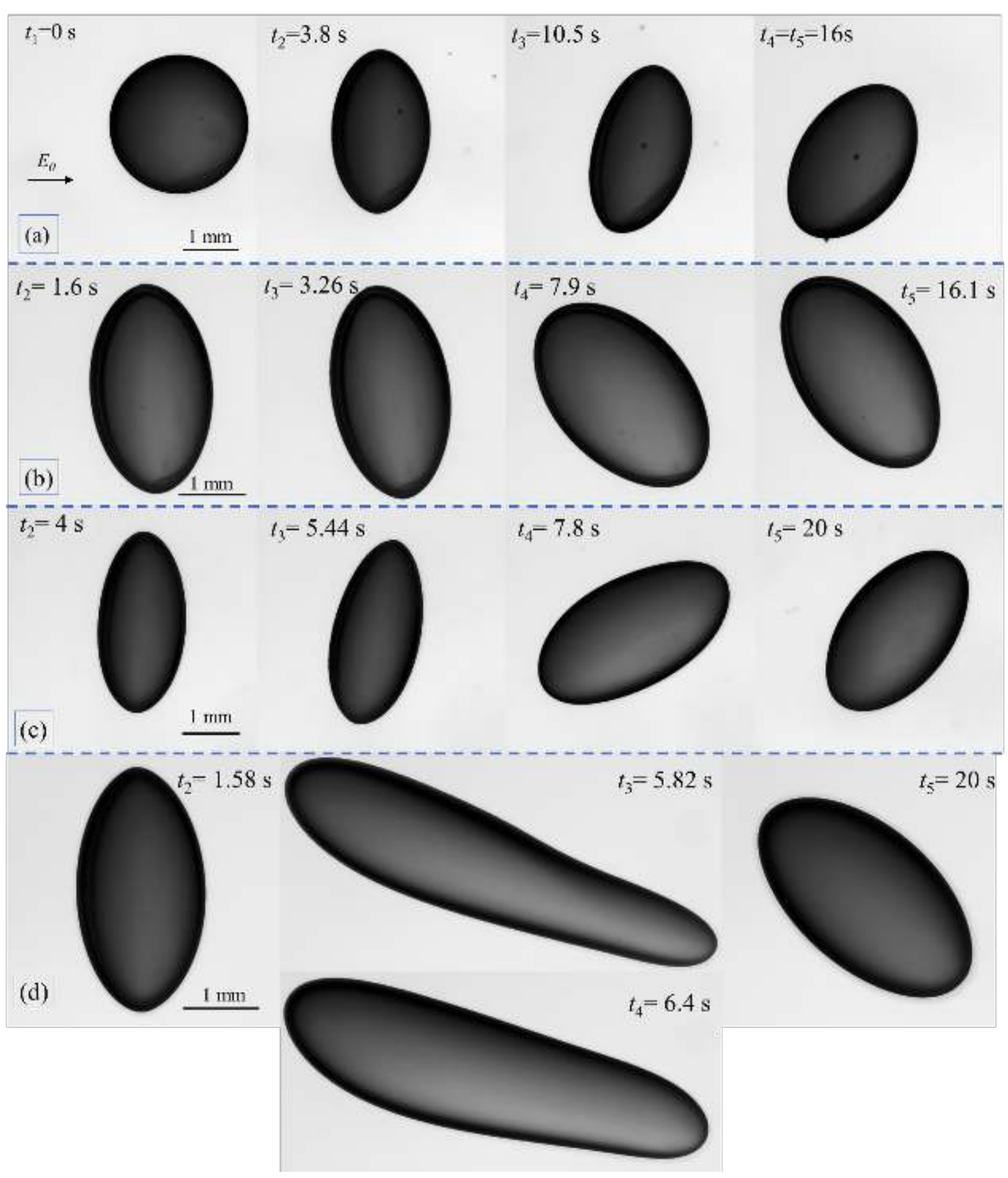

**Figure 2.** Experimentally captured images of a deformed droplet under different electric field strengths ($E_0$): (a) 4.50 kV/cm, (b) 4.75 kV/cm, (c) 5.25 kV/cm, and (d) 5.75 kV/cm

However, at higher electric fields, $E_0$>4.5 kV/cm the droplet response departs from this simple monotonic trend and displays an oscillatory (Fig. 2c and 2d) evolution of the aspect ratio before reaching equilibrium. The origin of this behavior lies in the finite charge relaxation time, when the droplet rotates rapidly, the interfacial charge distribution cannot instantaneously follow the rotating dipole, leading to a phase lag between the applied electrostatic torque and the hydrodynamic deformation response. This mismatch causes the system to overshoot its steady-state shape, resulting in a cyclic increase and decrease of the aspect ratio. The oscillations are progressively damped due to viscous dissipation in both the droplet and the suspending medium, and the system ultimately converges to a steady, less-deformed configuration. This oscillatory relaxation thus reflects the competition between three key mechanisms: (i) the strong, time-dependent electrostatic stresses generated by rapid Quincke rotation, (ii) the capillary restoring forces associated with interfacial tension, and (iii) the viscous damping that dissipates the rotational and interfacial flow energy, this is why the droplet doesn't keep tilting indefinitely but instead reaches a constant orientation. The eventual equilibrium state represents a balance among these effects, while the oscillatory transient is an inherent signature of the droplet's electrohydrodynamic dynamics at fields far above the rotation threshold.

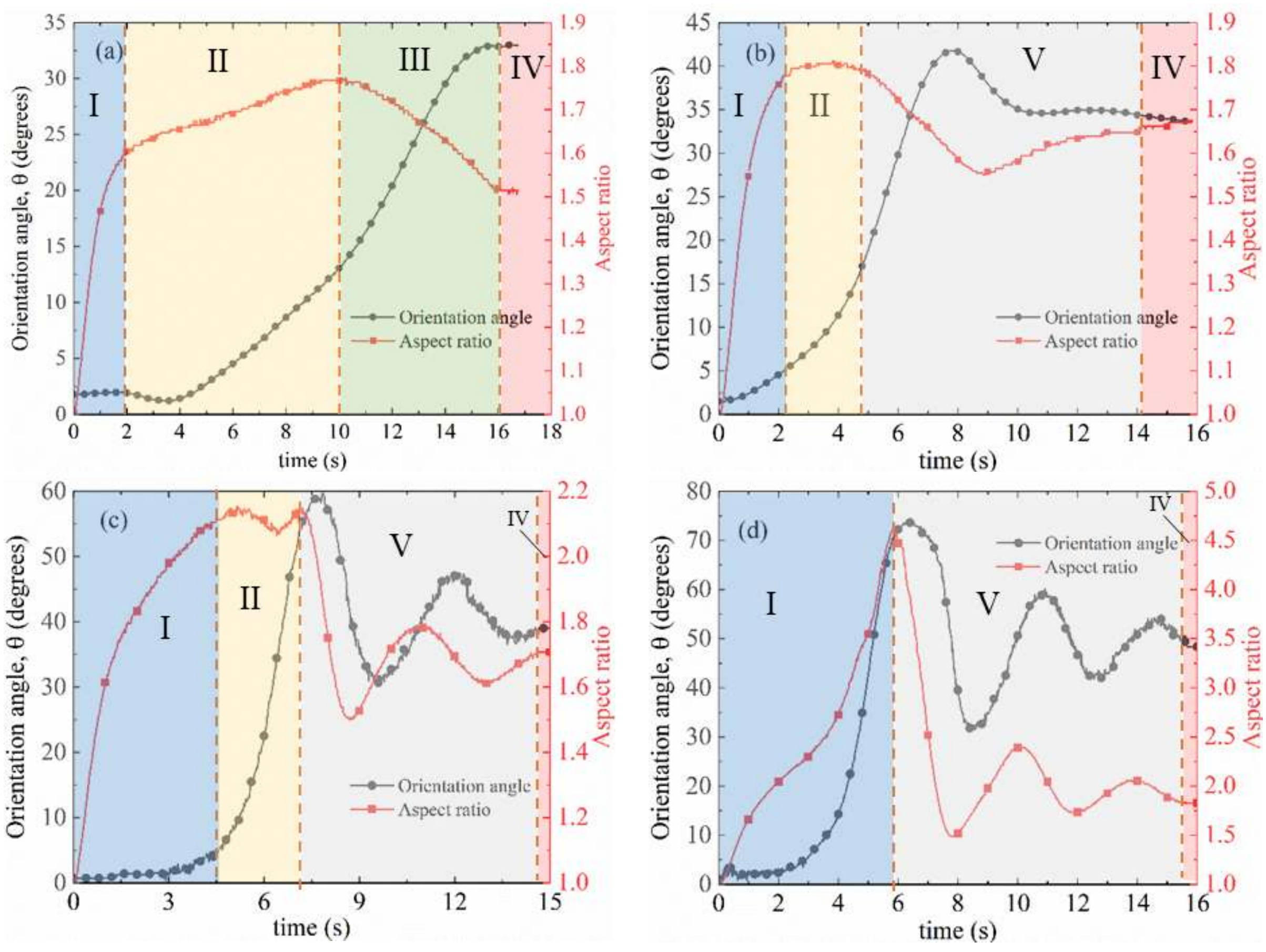

**Figure 3.** Variation of droplet orientation angle and aspect ratio at different electric field strengths ($E_0$): (a) 4.50 kV/cm, (b) 4.75 kV/cm, (c) 5.25 kV/cm, and (d) 5.75 kV/cm.

### *3D PTV (Shake the box Technique)*

Furthermore, this Study also investigates the deformation dynamic of the liquid droplet using the STB technique. Moreover, due to experimental constraints and practical considerations, a droplet diameter of 4.5 mm was selected for the present study employing the STB technique. However, the applied electric field strength was chosen such that the droplet behaviour transitioned from a steady-state deformation regime to a transient oscillatory deformation regime.

As discussed above, Figure 4 presents the images acquired during droplet deformation using three CMOS cameras (HiSense Zyla, Dantec). The figure demonstrates deformation dynamics that are qualitatively consistent with those observed in Figure 2(d). For the present case, the applied electric field strength was 3.75 kV/cm.

Figure 5 illustrates the flow circulation in the vicinity of a deforming droplet subjected to an applied electric field. The measurements are performed in the plane of the camera sensor, corresponding to the mid-plane of the initially undeformed droplet. As an initial response to the electric field, the droplet deforms into an oblate shape (Fig. 5a). This deformation is accompanied by the development of four distinct vortical structures, each occupying a separate quadrant around the droplet. The resulting flow pattern exhibits a quadrupolar organization that is consistent with the symmetry and geometry of the deformed droplet. In the flow vector field, the blue dots indicate the observed centers of flow circulation, which are used to identify the coherent flow structures.

At comparatively lower electric field strengths, such fourfold symmetric vortical structures remain sharp, stable, and well-defined, as previously reported by Karp et al. (2024). In contrast, under the present high electric field conditions, the droplet interface undergoes rapid and pronounced deformation, which strongly disrupts both the internal and external flow circulation. As a consequence of the continuously evolving droplet interface, the flow field does not maintain stable or persistent vortical structures. Instead, the circulation patterns undergo constant reorganization in response to the transient deformation dynamics. As shown in Fig. 5b, the initially quadrupolar flow structure transitions into two dominant coherent circulation zones as the droplet shape evolves. With further progression in time, the droplet enters a spinning motion characteristic of the Quincke rotation regime. During this stage, illustrated in Fig. 5d, two coherent circulation structures are again observed within the droplet interior.

This evolution of the circulation zones appears to be a distinctive feature of the oscillatory deformation dynamics and is consistent with the interfacial behavior observed in the shadowgraph imaging results presented earlier. Eventually, the transient deformation subsides and the droplet reaches a quasi-equilibrium state while undergoing sustained Quincke rotation as shown in Fig. 5e. During the deformation process, newly formed coherent structures emerge transiently and subsequently dissipate as the interface evolves, indicating a continuous adjustment between the flow field and the deforming boundary. Over the observation period, the initially observed multiple vortical structures progressively merge and reorganize into a single dominant circulation zone. This transition is indicative of a dynamical regime shift, ultimately consistent with the onset of Quincke rotation of the droplet.

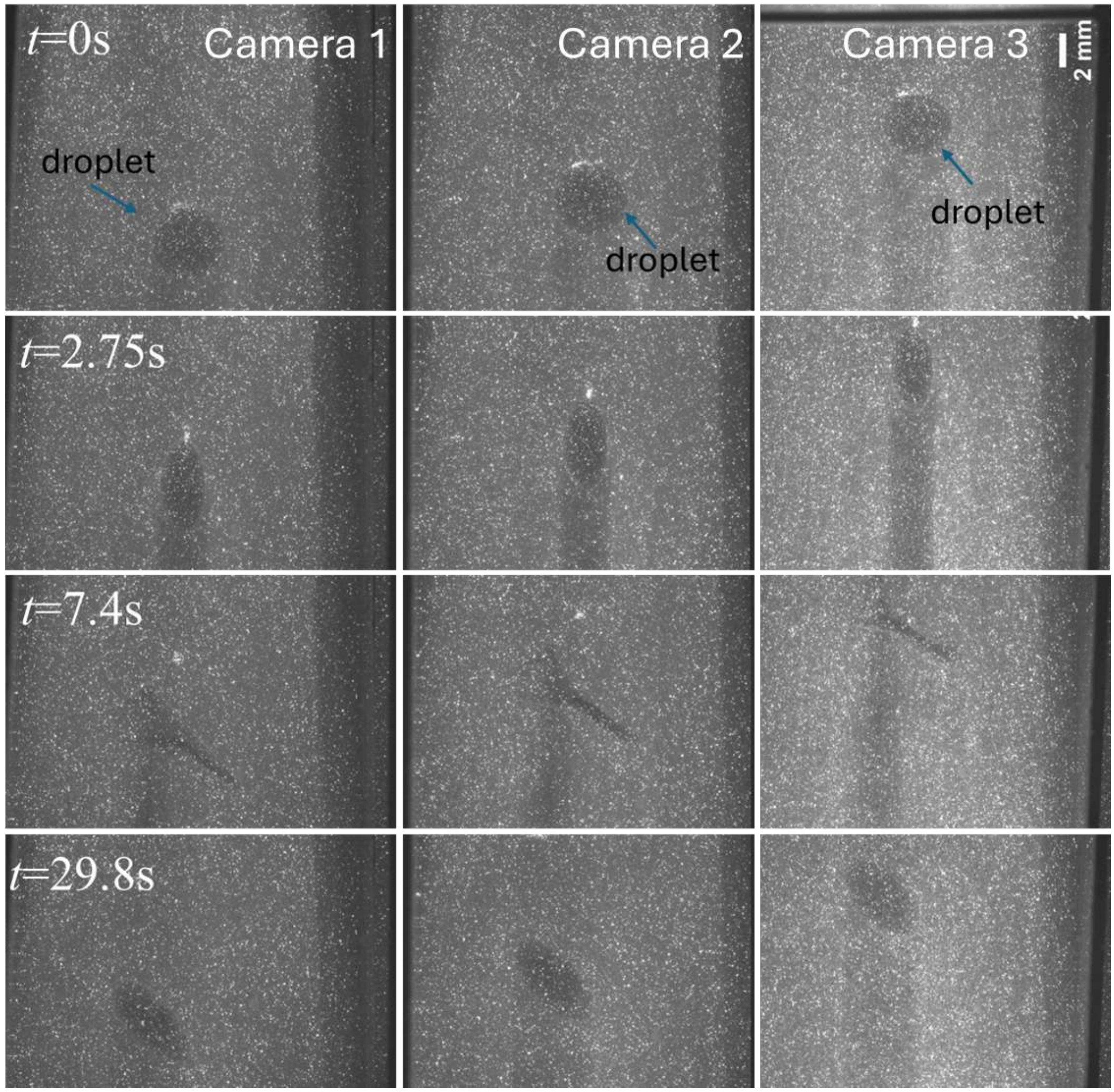

**Figure 4.** Sequence of droplet deformation dynamics captured in raw images from three synchronized cameras, showing seeded glass tracer particles under an applied electric field strength of 3.75 kV for a droplet with a diameter of 4.5 mm.

It is also important to note that the Shake-the-Box (STB) Lagrangian Particle Tracking technique has been employed here primarily as a preliminary demonstration to assess its suitability for investigating highly transient and oscillatory droplet deformation dynamics. The present study is primarily focused on the qualitative visualization of the rapidly evolving flow structures associated with strongly deforming droplet interface under the presence of an applied electric field. In this regime, the interfacial dynamics of the droplet are highly unsteady, which leads to continuous reorganization of both the internal and external flow fields. As a result, the present analysis is restricted to capturing the morphological evolution and the associated changes in flow topology, rather than providing a quantitative characterization of the underlying electrohydrodynamic mechanism.

A rigorous quantitative assessment of the observed flow transitions, including detailed measurements of velocity fields, circulation strength, and transition criteria between different dynamical regimes, is beyond the scope of the present work and will be addressed in a separate comprehensive study.

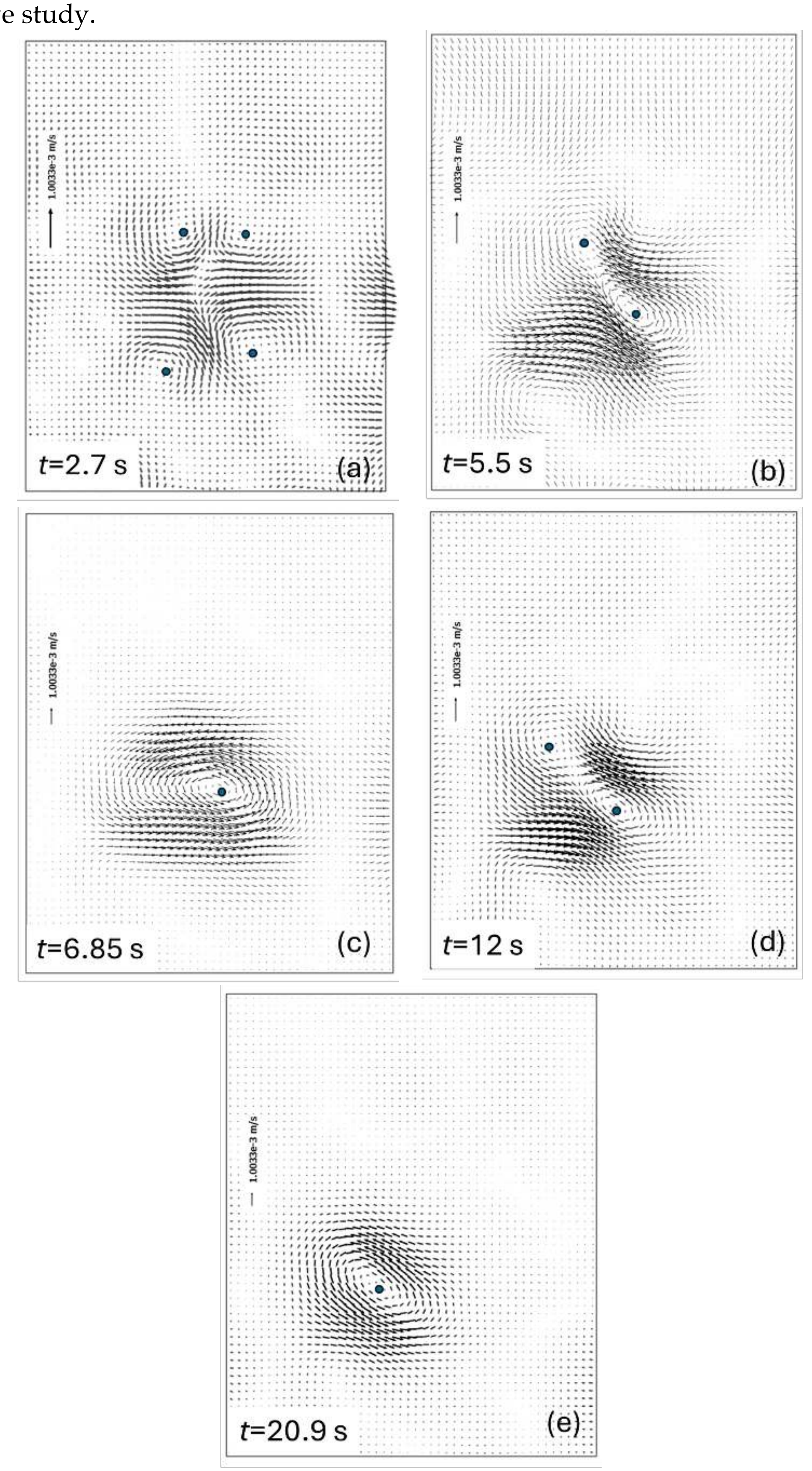

**Figure 5.** Temporal evolution of the flow vector field in the vicinity of the droplet.

## 4. Conclusions

In this study, an experimental investigation has been performed to analyze the deformation dynamics of a silicon oil droplet immersed in castor oil under varying electric field strengths. A comparative analysis of the temporal evolution of the droplet's aspect ratio and orientation angle reveals the distinct deformation regimes that are strongly dependent on the intensity of applied electric field. At lower field strengths, the droplet underwent steady, monotonic deformation accompanied by rotation or tilting along the direction of the electric field. In contrast, at higher field strengths, the deformation exhibited cyclic or oscillatory behaviour, with the aspect ratio oscillating in time and the orientation angle increasing significantly.

Furthermore, to evaluate the flow circulation both inside and outside the droplet, a three-dimensional particle tracking velocimetry (3D-PTV) technique, specifically the Shake-The-Box (STB) method, was employed. This approach was used to investigate the evolution of flow circulation in situ during the transient behaviour of the droplet. The 3D particle tracking results revealed a highly unsteady and developing flow-circulation pattern that varied significantly over time, highlighting the strongly transient nature of the flow field. However, the present 3D-PTV investigation is primarily qualitative and serves as a preliminary assessment of the applicability of the Shake-The-Box technique to the current experimental configuration. A more comprehensive and quantitative analysis will be the focus of future work, enabling a deeper understanding of the underlying flow dynamics.

These findings advance our understanding of the electrohydrodynamic flow inside the droplet in multiphase flow systems, providing new insights into the mechanisms governing deformation transitions and oscillatory dynamics under strong electric fields.

**Acknowledgments**
This project was financed by the Normandie region under the grant 22E05147. The measurements of the dielectric properties were conducted by Nicolas Delpouve and Laurent Delbreilh from the GPM (Groupe de Physique des Materiaux) of the UFR Sciences et Techniques de Rouen, whose contribution is also hereby acknowledged.